\documentclass[a4,12pt,fleqn]{article}
\usepackage{epsfig,times,latexsym,enumerate,lscape,float,flafter,amsmath}

\newcommand{\prd}{Phys. Rev. D } 
\newcommand{\mnras}{MNRAS }

\begin{document}

\def\nocropmarks{\vskip5pt\phantom{cropmarks}}
%

\title{Earth azimuth effect in the bank of search templates for an all sky search of the continuous gravitational wave}

\author{\footnotesize S.K. Sahay\thanks{Present address: BITS-Pilani, Goa campus, NH 17B, Bye pass road, Zuarinagar - 403726, Goa, India 
}\hspace{2.0mm}\thanks{E-mail: ssahay@bits-goa.ac.in}\\ The Wise Observatory, and the School of Physics and Astronomy, \\
Raymond and Beverly Sackler Faculty of Exact Sciences,\\
Tel Aviv University, Tel Aviv 69978,Israel\\ssahay@wise.tau.ac.il
}

\maketitle

\begin{abstract}
We study the problem of all sky search in reference to continuous 
gravitational wave (CGW) whose wave-form are known in advance. We employ  the concept 
of Fitting Factor and study the variation in the bank of search templates with different Earth azimuth at $t=0$.  We found that the number of 
search templates varies significantly. Hence, accordingly, 
the computational demand for the search may be reduced up to two orders by time shifting the data.
\noindent {\bf Keywords:} gravitational wave, data analysis, all sky search, LIGO.
\end{abstract}

\section{Introduction} 
 The ground based laser interferometric gravitational wave (GW)  
detectors viz. LIGO\cite{1}, VIRGO\cite{2}, GEO600\cite{3} 
TAMA 300\cite{4} produce a single data stream that may contain 
continuous, chirp, burst and stochastic GW. These detectors don't point, 
but rather sweep their broad 
quadrupolar beam pattern across the sky as the earth moves. Hence the data 
analysis system will have to carry
out all sky searches for its sources. In this, the search of continuous 
gravitational wave (CGW) without a priori knowledge appears to be
computationally quite demanding even by the standard computers
expected to be available in the near future. It appears that due to limited computational resource it will be not feasible to do all sky all frequency 
search for the CGW in the months/year data set. Hence for the search, one have to look for the signal in the short 
duration of data set. Understanding the problem and the target sensitivity of the advance LIGO, 
it may be 
feasible to do all sky search for one day data set at low and in narrow frequency band. The search may be more significant if it is done in the frequency 
band where most of the Pulsars are detected by other means. Also, the choice of optimal data 
processing and a clever programming is also
integral part of a solution to this problem. Amongst these the pre-correction
of time series due to the Doppler modulation before the data is processed may 
be a method, which will reduce the computational requirements. 
In reference to this, Schutz\cite{5} has introduced the concept of patch in the 
sky as the region of space throughout which the required Doppler correction 
remains the same. He shown that the number of patches required 
for $10^7$ sec. observation data set of one KHz signal would be about $1.3 
\times 10^{13}$. This also implies that the bank of search templates require for the match filtering in an 
all sky 
search. However, the size of the patch would also depend on the data analysis 
technique being employed, which in turn depend on the parameters 
contain in the phase of the modulated signal. Hence, in this paper after incorporating
the azimuth of the Earth ($\beta_{orb}$) at $t=0$ in the Fourier transform (FT) obtained by 
Srivastava and Sahay\cite{6}, we investigate 
its effect in the bank of search templates for one sidereal data set. 
Hence in the next section we extend the FT of the Frequency modulated (FM) CGW obtained by them\cite{6} with taking account of $\beta_{orb}$. In section 3 we employ the  concept of Fitting Factor (FF)\cite{7} and check the 
cross correlation of the templates with the corresponding data set exceeds the
preassigned threshold by considering the source location as parameters at 
different $\beta_{orb}$ of the signal manifold and compute the number of 
search templates require for an all sky search applicable to such 
analysis. We present our conclusion in section 4.

 . 
\section{Fourier transform of the Frequency modulated continuous gravitational wave}

The FT analysis of the Frequency modulated CGW has been done by 
Srivastava and Sahay\cite{6} by taking account the effects arising due to 
the rotational as well 
as orbital motion of the Earth. However, they have neglected an important 
parameter 
$\beta_{orb}$. To obtain FT by taking account of $\beta_{orb}$, we rewrite the phase of the received CGW signal of frequency $f_o$ at 
time $t$ with some modification given by them\cite{6} and may be written as

\begin{eqnarray}
 \Phi (t) & = & 2\pi f_o t + {\cal Z}\cos (a\xi_{rot} - \sigma ) + 
{\cal N}\cos (\xi_{rot} - \delta ) - {\cal M}
\end{eqnarray}

\noindent where \\
\begin{equation}
\left.\begin{array}{lcl}
\vspace{0.2cm}
{\cal M}& = & \frac{2\pi f_o}{c}\left(R_{se}\sin\theta\cos\sigma + \sqrt{{\cal P}^2 + {\cal Q}^2 }\cos\delta\right) , \\
\vspace{0.2cm}
{\cal Z}& = & \frac{2\pi f_o}{c} R_{se}\sin\theta\, , \quad {\cal N}\; = \; \frac{2\pi f_o}{c} \sqrt{{\cal P}^2 + {\cal Q}^2 } \, ,\\
\vspace{0.2cm}
\vspace{0.2cm}
{\cal P}& = & R_e\sin\alpha (\sin\theta\sin\phi\cos\epsilon  + \cos\theta \sin\epsilon )\, ,\\
\vspace{0.2cm}
{\cal Q}& = & R_e \sin\alpha\sin\theta\cos\phi \, ,\\
\vspace{0.2cm}
\sigma &=& \phi - \beta_{orb}\, , \quad \delta = \tan^{- 1}\frac{{\cal P}}{{\cal Q}} - \beta_{rot}\, , \\
\vspace{0.2cm}
a &= & w_{orb}/w_{rot}\; \approx \; 1/365.26, \quad w_{orb}t \;  = \; a\xi_{rot} , \\
\vspace{0.2cm}
{\bf\textstyle n} & = & \left(\sin\theta\cos\phi , \; \sin\theta\sin\phi\, , \;\cos\theta\right) , \quad \xi_{rot}\;  = \; w_{rot}t \\
\end{array} \right\}
\end{equation}

\noindent  where $\theta$, $\phi$, $R_{e}$, $R_{se}$, $w_{orb}$ $w_{rot}$, $\alpha$ and 
$\epsilon$ represent 
respectively the celestial co-latitude, longitude, Earth radius, average distance between Earth centre from 
the origin of SSB frame, orbital and rotational angular velocity of the 
Earth, co-latitude of the detector and obliquity of the ecliptic. Here $\beta_{orb}$ and $\beta_{rot}$ are the azimuth of 
the Earth and detector at $t=0$ respectively.\\ 

\noindent  Now, the two 
polarisation functions of CGW can be described as
\begin{equation}
h_+(t) = h_{o_+}\cos [\Phi (t)] 
\label{eq:hpt}
\end{equation}
\begin{equation}
h_\times (t) = h_{o_\times}\sin [\Phi (t)]
\label{eq:hct}
\end{equation}

\noindent where $h_{o_+}$, $h_{o_\times}$ are constant amplitude of the two 
polarizations.\\ 

\noindent Considering the function 
\begin{equation}
h(t) = \cos[\Phi (t)]
\label{eq:cosphit}
\end{equation}

\noindent the FT for one sidereal day may be given as 

\begin{equation}
\left[\tilde{h}(f)\right]_d = \int_0^{T_{obs}} \cos[\Phi (t)]e^{-i2\pi ft}dt\; ; \qquad
T_{obs} = one\; sidereal\; day = 86164\; s
\label{eq:hf1}
\end{equation}
\noindent which may be split into 
\begin{equation} \left[\tilde{h}(f)\right]_d = I_{\nu_-} + I_{\nu_+} \; ;
 \label{eq:inu}
 \end{equation}
\begin{eqnarray}
\label{eq:dayinu}
 I_{\nu_-}& =& {1\over 2 w_{rot}}\int_0^{2\pi} e^{i \left[\xi\nu_- +{\cal Z}\cos (a\xi - \sigma ) + {\cal N}\cos (\xi - \delta ) -{\cal M}  \right] } d\xi\, ,\\
I_{\nu_+}& = &  {1\over 2 w_{rot}}\int_0^{2\pi} e^{- i \left[\xi\nu_+ + {\cal Z} \cos (a\xi - \sigma ) + {\cal N}\cos (\xi - \delta ) - {\cal M} \right] } d\xi\, , \\
\nu_{\pm}& = & \frac{f_o \pm f}{f_{rot}} ; \quad \xi \; = \; \xi_{rot} \; = \;
w_{rot}t 
\end{eqnarray}

\noindent As $I_{\nu_+}$ contributes very little to $\left[\tilde{h}(f)\right]_d$. Hence, hereafter, we drop $I_{\nu_+}$ 
from Eq.~(\ref{eq:inu}) and write $\nu$ in place of $\nu_-$. Using 
the identity

\begin{equation}
e^{\pm i\kappa\cos\vartheta} = J_o (\pm\kappa) + 2 \sum_{l = 1}^{l = 
\infty} i^l J_l (\pm\kappa)\cos l\vartheta
\label{eq:bessel}
\end{equation}

\noindent we obtain
\begin{eqnarray}
\left[\tilde{h}(f)\right]_d &\simeq & \frac{1}{2 w_{rot}} e^{- i {\cal M}} 
\int_0^{2\pi} e^{i\nu\xi} \left[ J_o( {\cal Z} ) + 2 
\sum_{k = 1}^{k =  \infty} J_k ({\cal Z}) i^k \cos k (a\xi - \sigma )\right] 
 \nonumber \\
&& \times\,\left[J_o( {\cal N} ) + 2 \sum_{m = 1}^{m =  \infty} J_m ({\cal N}) i^m 
\cos m (\xi - \delta )\right] d\xi
\end{eqnarray}

\noindent where $J$ stands for the Bessel function of the first kind. After 
integration we get
\begin{equation}
\label{eq:hfd}
\left[\tilde{h}(f)\right]_d  \simeq  \frac{\nu}{2 w_{rot}} \sum_{k  =  - 
\infty}^{k = \infty} \sum_{m = - \infty}^{m =  \infty} e^{ i {\cal A}}{\cal 
B}[ {\cal C} - i{\cal D} ] \; ; \;
\end{equation}  
\begin{equation}
\left.\begin{array}{lcl}
\vspace{0.2cm}
{\cal A}&  = &{(k + m)\pi\over 2} - {\cal M}  \nonumber \\
\vspace{0.2cm}
{\cal B} & = & {J_k({\cal Z}) J_m({\cal N})\over {\nu^2 - (a k + m)^2}} \nonumber \\
\vspace{0.2cm}
{\cal C} &= & \sin 2\nu\pi \cos ( 2 a k \pi - k \sigma - m \delta ) - 
{ a k + m \over \nu}\{\cos 2 \nu \pi \sin ( 2 a k \pi - k \sigma - m \delta )
+ \sin ( k \sigma + m \delta )\}\nonumber \\
\vspace{0.2cm}
 {\cal D} & = & \cos 2\nu\pi \cos ( 2 a k \pi - k \sigma - m \delta ) + 
 {k a +m \over \nu}\sin 2 \nu \pi \sin ( 2 a k \pi - k \sigma - m \delta )
 - \cos ( k \sigma + m \delta )  \nonumber 
\end{array} \right\}
\end{equation}


\noindent Now its straight forward to obtain the FT of the two polarisation states of the wave and can be written as
\begin{eqnarray}
\label{eq:hfpd}
\left[\tilde{h}_+(f)\right]_d&=&h_{o_+}\left[\tilde{h}(f)\right]_d \nonumber \\
&\simeq & \frac{\nu h_{o_+}}{2 w_{rot}} \sum_{k  =  - \infty}^{k =  
\infty} \sum_{m = - \infty}^{m =  \infty} e^{ i {\cal A}}{\cal B}[ {\cal C} - i{\cal D} ] \; ;
\end{eqnarray}

\begin{eqnarray}
\label{eq:hfcd}
\left[\tilde{h}_\times (f)\right]_d &=&- i h_{o_\times}\left[\tilde{h}(f)\right]_d \nonumber \\
&\simeq & \frac{\nu h_{o_\times}}{2 w_{rot}} \sum_{k  =  - \infty}^{k =  
\infty} \sum_{m = - \infty}^{m =  \infty} e^{ i {\cal A}}{\cal B}[ {\cal D} - i{\cal C} ] 
\end{eqnarray}

\noindent For the analysis we reduce the computing time $\approx 50\%$ by using 
symmetrical property of the Bessel functions by rewriting the Eq.~(\ref{eq:hfd}) as 

\begin{eqnarray}
\left[\tilde{h}(f)\right]_d&\simeq & { \nu \over w_{rot}}\left[ {J_o({\cal Z}) J_o({\cal N}) \over 2\nu^2}\left[ \{ \sin{\cal M} - \sin
({\cal M} - 2 \nu\pi )\}\; +  i\{ \cos{\cal M} - \cos ({\cal M} - 2 \nu
\pi )\} \right]\; \right.+ \nonumber \\
&& J_o ({\cal Z})\sum_{m = 1}^{m = \infty} {J_m({\cal N})\over 
\nu^2 - m^2} \left[ ( {\cal Y} {\cal U} -  {\cal X} {\cal V} ) - i ( 
{\cal X} {\cal U} + {\cal Y} {\cal V} ) \right]\; + \nonumber \\ 
&& \left.\sum_{k = 1 }^{k = \infty}\sum_{m = -
\infty}^{m = \infty} e^{ i {\cal A}}{\cal B}\left(
\tilde{{\cal C}} - i\tilde{{\cal D}} \right)\right]\; ;
\label{eq:fm_code}
\end{eqnarray} 

\begin{equation}
\left.\begin{array}{ccl}
{\cal X}& =& \sin ({\cal M}  - m \pi/2 )\\
{\cal Y}& =& \cos ({\cal M} - m \pi/2 )\\
{\cal U}& =& \sin 2\nu\pi \cos m ( 2\pi - \delta ) - {m\over \nu}\left\{\cos
2\nu\pi \sin m (2\nu\pi - \delta ) - \sin m\delta\right\}\\
{\cal V}& =& \cos 2\nu\pi \cos m (2\pi - \delta ) + {m\over \nu}\sin 
2\nu\pi \sin m ( 2\pi - \delta ) - \cos m\delta\\
\end{array}\right\}
\end{equation}

\section{Bank of search templates for an all sky search} 
\begin{figure}
\centering\epsfig{file=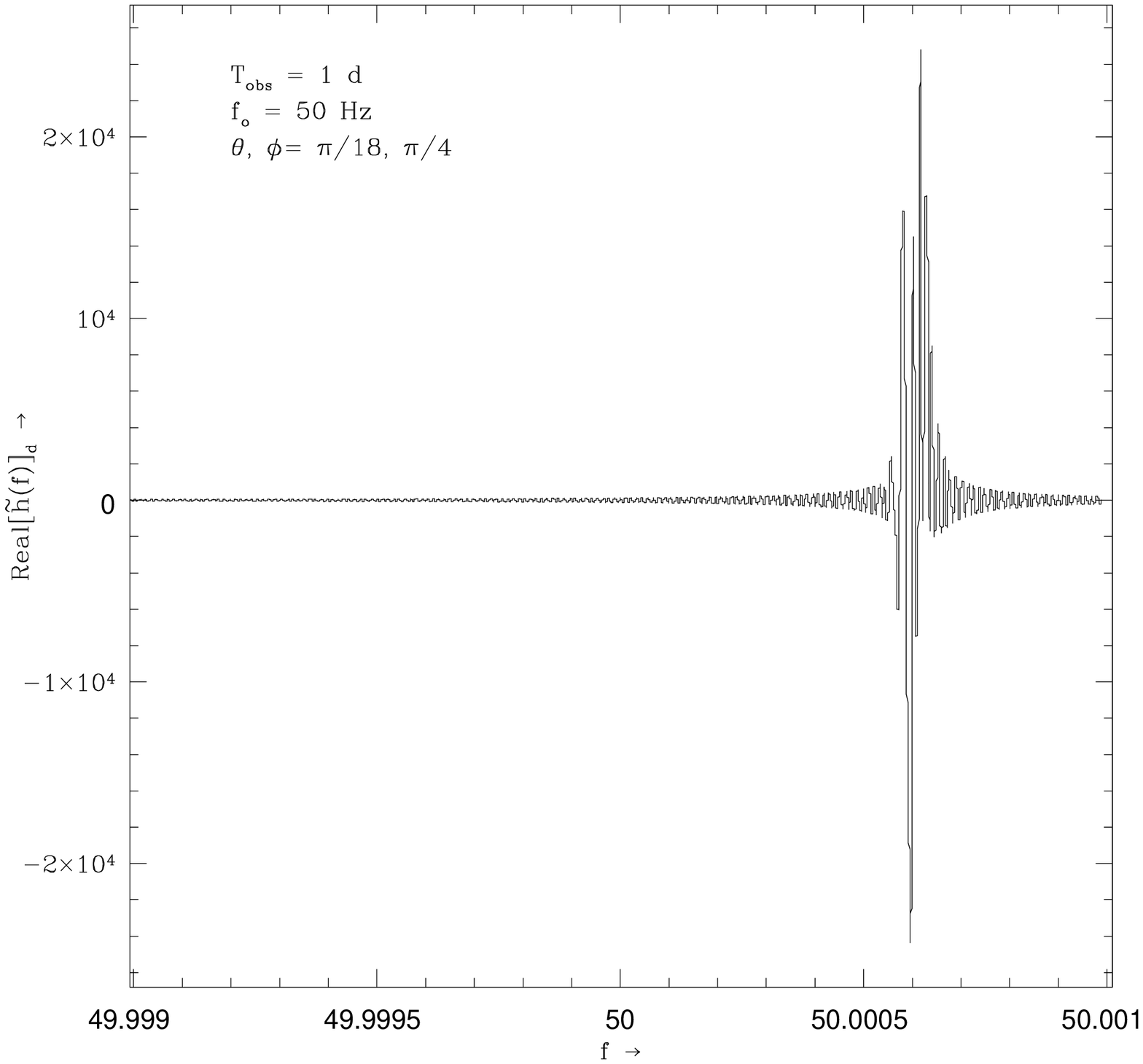,height=5.5cm}
\caption{Spectrum of a Doppler modulated signal when $\beta_{orb} = 0$.}
\label{fig:orb0}
\end{figure}

\begin{figure}
\centering\epsfig{file=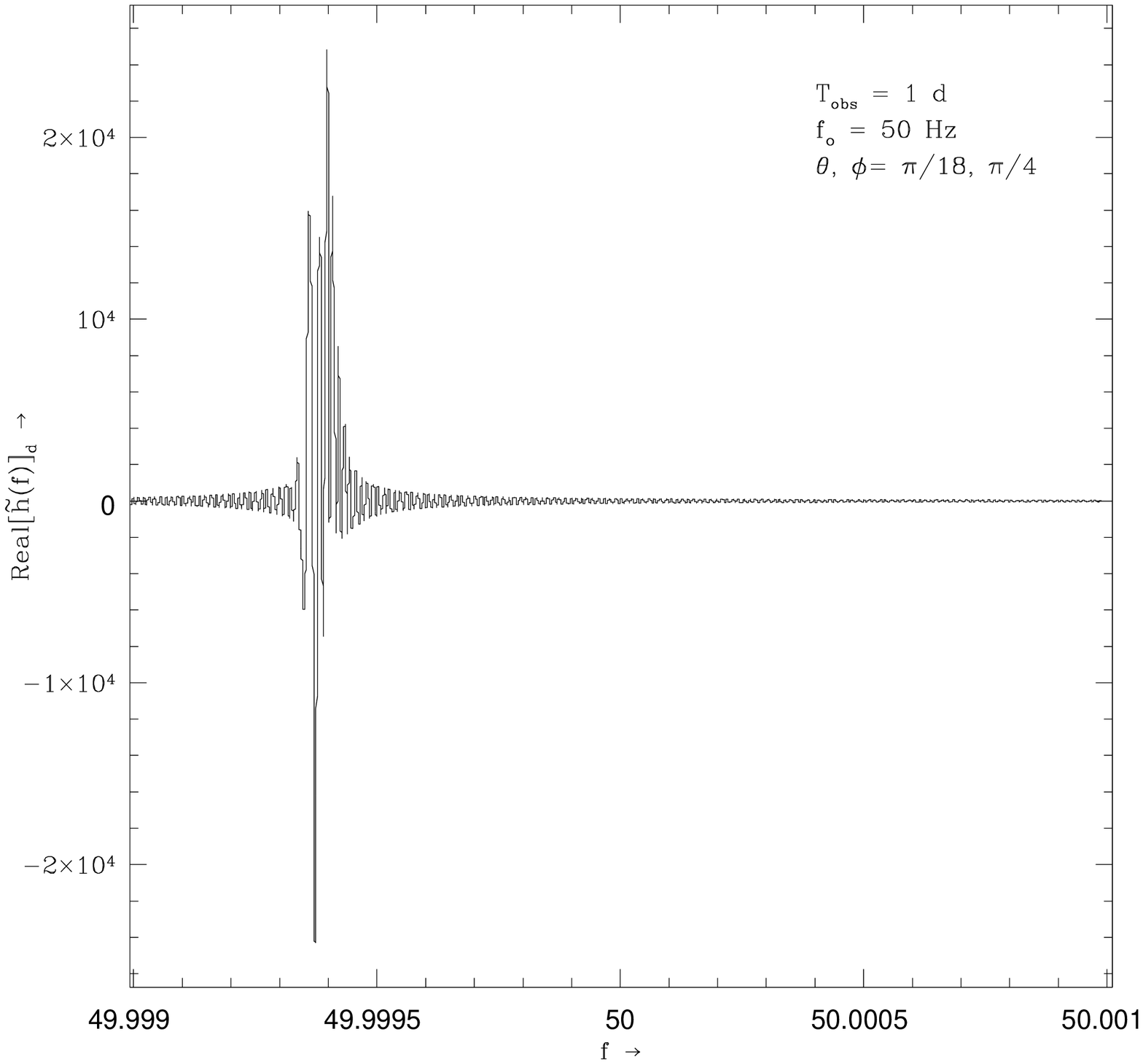,height=5.5cm}
\caption{Spectrum of a Doppler modulated signal when $\beta_{orb} = \pi /2$.}
\label{fig:orb90}
\end{figure}

The study of templates has been made by many research workers\cite{8,9,10,11}. However,
the question of possible minimum efficient interpolated representation
of the correlators for an all sky search is a problem of interest. 
In this, the study of the variation in the bank of search templates for the  short duration 
of data set in reference to the 
parameter contain in the phase of modulated signal is very important. Hence we check 
the  
 effect of $\beta_{orb}$ in the modulated signal by plotting $ \left[\tilde{h}(f)\right]_d$ (Figures~(\ref{fig:orb0}) and~(\ref{fig:orb90})) for LIGO detector at Hanford (the position and orientation of the detector 
can be found in Ref. 12) of unit amplitude signal for
\begin{equation}
\label{eq:dayloc}
\left.\begin{array}{ll}
\vspace{0.2cm}
f_o = 50\;  Hz\, , & \beta_o = 0, \pi/2\\
\vspace{0.2cm}
\theta = \pi /18\, , & \phi = \pi /4 \\
\end{array} \right\}
\end{equation}

Here we take the ranges\cite{13,14} of $k$ and $m$ as 1 to 27300 and -10 
to 10 respectively because the value of Bessel functions decreases rapidly as 
its order exceed the argument. From the plot, we observe the obvious but major 
shift in the spectrum  
with $\beta_{orb}$. Understanding the effect  we check the variation in the bank of 
search templates at different $\beta_{orb}$. 

\par The bank of search 
templates are discrete set of signals from
among the continuum of possible signals. Consequently all the 
signals will not get detected with equal efficiency. However, it is possible to 
choose judiciously the set of templates so that all the signals of a given amplitude 
are
detected with a given minimum detection loss. FF is one of the  standard 
measure for deciding what class of wave form is good enough and 
quantitatively describes the
 closeness of the true signals to the
template manifold in terms of the reduction of SNR arising due to the cross 
correlation of 
a signal outside the manifold with the best matching templates lying inside the 
manifold, given as

\begin{eqnarray} 
{FF} & = & \frac{\langle h(f)|
h_T(f;\theta_T , \phi_T)\rangle}{\sqrt{\langle h_T(f;\theta_T , \phi_T )|h_T(f;
\theta_T , \phi_T )\rangle\langle h(f)|h(f)\rangle}}
\label{eq:ff1}
\end{eqnarray}

\noindent where $h(f)$ and $h_T(f; \theta_T , \phi_T)$ represent respectively the
FTs of the actual signal wave form and the templates. The inner product of two 
waveform $h_1$ and $h_2$ is defined as
\begin{eqnarray}
\langle h_1|h_2\rangle & =& 2\int_0^\infty \frac{\tilde{h}_1^*(f)\tilde{h}_2(f)
+ \tilde{h}_1(f)\tilde{h}_2^*(f)}{S_n(f)}df \nonumber \\
 & = &
4\int_0^\infty \frac{\tilde{h}_1^*(f)\tilde{h}_2(f)}{S_n(f)}df
\label{eq:ip}
\end{eqnarray}

\noindent where $^*$ denotes complex conjugation, $\tilde{}$ denotes the FT of the quantity 
underneath. $S_n(f)$ is the 
spectral noise density of the detector and has been taken stationary and Gaussian as the bandwidth of the signal is extremely narrow.  

\begin{figure}
\centering\epsfig{file=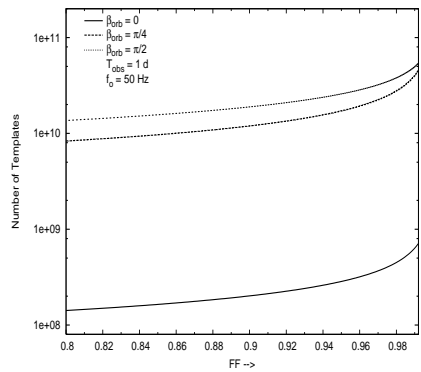,height=5.0cm,width=5.7cm}
\caption{Variation in the number of search templates with FF at at different $\beta_{orb}$.}
\label{fig:templates}
\end{figure}

To compute the number of search templates we consider the LIGO detector at Hanford, receive a CGW signal of 
frequency $f_o=50$ Hz from a source 
located at $(\theta , \phi) = (0.1^o,30^o)$. We chosen the data set such that $\beta_{orb} =0$ at $t=0$. In this case we take
the ranges of $k$ and 
$m$ as $1$ to $310$ and $-10$ to $10$ respectively and bandwidth equal to  
$2.0 \times 10^{-3}$ Hz for the integration. Now, we select 
the spacing $\bigtriangleup\theta = 4.5 \times 10^{-5}$, thereafter we maximize over $\phi$ by introducing 
spacing $\bigtriangleup\phi$ in the so obtained bank of search templates and determine 
the resulting $FF$. In similar manner we obtain the $FF$ at $\beta_{orb}=\pi/4$ and $\pi/2$. The results obtained are shown in the Fig.~(\ref{fig:templates}). We observe that the nature of the curve  is
similar. Hence, it may be interesting to obtain a best fit of the graphs and may be given as

\begin{equation}
N_{Templates} = 10^{15} [c_o +c_1x -c_2x^2 +c_3x^3 -c_4x^4 +c_5x^5 -c_6x^6 +c_7x^7]\; ;
\end{equation}

$\qquad 0.80 \le x \le 0.995$\\
where $c_o, c_1, c_2, c_3, c_4, c_5, c_6, c_7$ are the constants given 
in the Table~(\ref{table:coefficients}).

\begin{table}
\caption{Coefficients of the best fit graphs obtained for the 
bank of search templates.}
{\begin{tabular}{ccccccccc}\\
\hline
&&&&&&\\
$\beta_{orb}$ & $c_o$ & $c_1$ & $c_2$ & $c_3$ & $c_4$ & $c_5$ & $c_6$ & $c_7$ \\
&&&&&&\\ \hline
&&&&&&\\
$0^o$&$-4.36537$& $34.4523 $ & $116.44$ & $218.462 $&$245.734$&$165.719$&$62.0404 $&$9.94642$\\
&&&&&&\\
$45^o$&$-403.012$& $3187.44$ & $10795.5$ & $20296.6$&$22877.7 $&$15459.9$&$5799.45$&$931.641$\\
&&&&&&\\ 
$90^o$ &$-263.622$& $2086.48$ & $7071.80$ & $13305.6 $&$15008.9 $&$10150.3$&$3810.65$&$612.645$\\ 
&&&&&&\\\hline
\end{tabular}}
\label{table:coefficients}
\end{table}

In view of the above investigation, the grid
spacing ($\bigtriangleup \theta, \bigtriangleup\phi $) in the ($\theta,\phi)-$parameter of templates may be
expressed as   
\begin{equation}
\bigtriangleup\theta = {\cal F}(FF, f_o, \theta ,\phi ,T_{obs}, \beta_{orb} )
\label{eq:thetafn}
\end{equation}
\noindent Similarly,
\begin{equation}
\bigtriangleup\phi = {\cal G}(FF, f_o, \theta ,\phi ,T_{obs}, \beta_{orb})
\label{eq:phifn}
\end{equation}

In Ref. to 15, for one sidereal data set, the dependence of {\it FF \/} on the 
template variables $\theta_T$ and $\phi_T$ is given as
\begin{equation}
FF = e^{- 0.00788(\theta \; - \;\theta_T)^2}
\label{eq:thetaff}
\end{equation}
\begin{equation}
FF = e^{- 0.01778(\phi\; - \;\phi_T)^2}
\label{eq:phiff}
\end{equation}

\noindent From Eqs.~(\ref{eq:thetafn}),~(\ref{eq:phifn}),~(\ref{eq:thetaff}) and~(\ref{eq:phiff}), we may write
\begin{equation}
{\cal F}(FF, 50, 0.1^o , 30^o ,1 d, \beta_{orb}) = \left[ - (0.00788)^{-1}\ln (FF)\right]^{1/2} 
\end{equation}

\begin{equation}
{\cal G}(FF, 50, 0.1^o , 30^o ,1 d, \beta_{orb}) = \left[ - (0.01778)^{-1}\ln (FF)\right]^{1/2} 
\end{equation}

\noindent Hence, for the selected $FF$ one can determine $\bigtriangleup\theta$ and
$\bigtriangleup\phi$. However,  there is no unique choice for it. 
Here we are interested in the assignment of
$\bigtriangleup\theta$ and  $\bigtriangleup\phi$ such that the 
spacing is maximum resulting into the least number of
templates. As we have mentioned earlier,  
there is stringent requirement on reducing computer time. Accordingly, 
there is serious
need of adopting some procedure/formalism to achieve this. For example, one
may adopt the method of
hierarchical search. 

\section{Conclusions}
We have incorporated the parameter $\beta_{orb}$ in the FM signal and investigated its effect in the bank of search templates for an all sky 
search. The analysis for complete response of the detector has not been done as the requisite 
is analogous to what presented by them\cite{6} and also not required for 
the templates analysis.

\par We observe that the change of $\beta_{orb}$  
 affects the spectra severely. Consequently in the bank search templates. 
On investigation for one sidereal day data set of a signal of $50$ Hz, we found that the number of search templates 
varies significantly with $\beta_{orb}$. For the case investigated we found that  for $FF=0.97$ approximately $24.8267 \times 10^{10}$, $22.7840 \times 10^{12}$ and $32.3097 \times 10^{12}$ search templates may be require, 
if $\beta_{orb} = 0, \pi/4$ and $\pi/2$ respectively. Hence, the 
computational demand may be reduce up to two orders by time 
shifting the data. One may optimize the require number of search templates 
 by doing analysis with the method given by Owen\cite{16}.

 \section*{Acknowledgment}
I am thankful to Prof. D.C. Srivastava, Department of Physics, DDU Gorakhpur University, Gorakhpur for useful discussions.

\end{document}